\documentclass[amsmath,amsfonts,amssymb,showpacs]{revtex4}
\usepackage{graphicx}
\def\beq{\begin{equation}}
\def\eeq{\end{equation}}
\def\bea{\begin{eqnarray}}
\def\eea{\end{eqnarray}}

\begin{document}

\title{Entanglement Generation by Electric Field Background}

\author{Zahra Ebadi}
\email{z.ebadi@ph.iut.ac.ir}
\author{Behrouz Mirza}
\email{b.mirza@cc.iut.ac.ir}

\affiliation{Department of Physics, Isfahan University of Technology, Isfahan 84156-83111, Iran}

\begin{abstract}
The quantum vacuum is unstable under the influence of an external electric field and decays into pairs of charged particles, a process which is known as the Schwinger pair production. We propose and demonstrate that this electric field can generate  entanglement. Using the Schwinger pair production for constant and pulsed electric fields, we study  entanglement  for scalar particles with zero spins and Dirac fermions.  One can observe the variation of  the entanglement produced for bosonic and fermionic modes with respect to different parameters.

\pacs{03.67.Bg, 03.67.-a, 04.62.+v}
\end{abstract}

\maketitle

\section{Introduction}
The quantum information theory is  important for the multitude of its promising new applications in such varied fields as quantum communication and teleportation, quantum cryptography, quantum computing, etc. The concept of entanglement also plays crucial roles in black hole thermodynamics \cite{muko, levay} and in the information loss problem \cite{horo, ahn, adesso}, which have given rise to
  many studies aimed at  measuring the generation and degradation of entanglement in a wide spectrum of  systems. These studies include investigation of entanglement in both inertial \cite{alsing} and non-inertial frames \cite{alsingg, funtess, alsinggg, martin, telee, bruschi} as well as its generation in expanding spacetime \cite{funtes2pla, funtesprd} and in relativistic quantum fields \cite{4funtes}.


  Although many of these works are far from being experimental, they are valuable as they offer a refined understanding of quantum information. In this paper, we explore the generation of entanglement using Schwinger pair production.
   For this purpose, we will investigate the effect of background electric field on the generation of entanglement
   for scalar and spinor fields.

   It is well known that when an external electric field is applied to the quantum electrodynamical vacuum, the vacuum becomes unstable and decays into pairs of charged particles. In fact, the quantum vacuum is unstable under the influence of an external electric field, as the virtual electron-positron dipole pairs  gain energy from the external field. When the field is sufficiently strong, these virtual pair particles gain the threshold pair creation energy and become real pairs. This remarkable phenomenon was first predicted by F. Sauter \cite{sauter} to be later refined by W. Heisenberg and H. Euler \cite{heisenberg} and formalized in the language of QED by Schwinger \cite{Schwinger1}, hence its designation nowadays as the  Schwinger pair production effect. This phenomenon has been investigated by scholars and workers from a variety of fields \cite{niki, dune}. Efforts in the 1900s and early 21st century aimed at descriptions of more realistic field configurations led to the development of different formalisms such as the quantum kinetic approach which were used for the  numerical computation of the Schwinger effect \cite{blasch}. Other approaches used include the closely related scattering-like formalism in terms of the Riccati equation \cite{dumlu}, the Dirac-Heisenberg-Wigner formalism \cite{heben}, and the numerical worldline formalism\cite{gies}. The critical electric field required for pair creation is almost $10^{16}V/cm$  which is too enormous to be directly observed. However, the feasibility of its  experimental realization in ultra-intense laser field system \cite{taji, dunne} has recently led  to a re-thinking of the Schwinger effect. It has been realized that the Schwinger limit laser intensity of $4\times10^{29}W/cm^{2}$  is not necessarily a strict limit and might be lowered by several orders of magnitude through manipulating the form of the laser pulses \cite{rsch, gvd, piaza, bulanov, monin}. Furthermore,  it has been proposed that the Schwinger pair production effect may be observed in graphene \cite{graphene}. These considerations motivated the authors to study the generation of entanglement using an electric field.

 The present paper is organized as follows. In Section II, we utilize the Schwinger effect for scalar particles with zero spin and Dirac fermions in the presence of a constant electric field. We will demonstrate that a constant electric field can generate the entanglement that its value can be determined. We will also consider the variation of the entanglement produced for bosonic and fermionic modes with respect to different parameters. In Section III,  we extend our investigation to the pulsed electric field.  Finally, conclusions will be presented in Section  VI.

\section{  Entanglement Generation IN A Constant Electric Field  }

The Minkowski vacuum becomes unstable by a strong electric field and decays into pairs of charged particles.
 One can use the '$in$' and '$out$' formalism in order to investigate the entanglement generation. '$In$' and '$out$' are related to asymptotic times $t= - \infty$ and $t= +\infty$, respectively. If the separable '$in$' state can be expanded in terms of the entangled '$out$' state, the generated entanglement can then be determined. The state of two particles $A$ and $B$ is a vector in a $(d\times d')$-dimensional Hilbert space $H_{ab}=H_{a}\otimes H_{b}$. The space $H_{ab}$ is the tensor product of the subspaces $H_{a}$ and $H_{b}$ of each particle. An element of the space $H_{ab}$ is written as $|\Phi\rangle_{ab}=\sum_{i,j} C_{ij}|i\rangle_{a}\otimes |j\rangle_{b}$. A state $|\Phi\rangle_{ab}\in H_{a}\otimes H_{b}$ is separable if $|\Phi\rangle_{ab}=|\Phi\rangle_{a} \otimes \Phi\rangle_{b}$. An entangled state is a state that is not separable \cite{shi}.

In the following subsections, we study entanglement entropy for charged scalar and fermion particles in the presence of an electric field.

\subsection{ Entanglement Entropy For Scalar Particles }

In the study of entanglement generation, we use asymptotic solutions of equation of motion for charged scalar particles in the presence of an electric field.

Consider an electric field along the z-direction. It is related to the gauge potential through $E_{z}(t)=-\partial A_{z}(t)/\partial t$. For a scalar particle of mass $m$ and Charge $q$, the Klein-Gordon equation on the four dimensional Minkowski spacetime with the metric $(+,-,-,-)$ is given by
\bea
 [(\partial_{\mu}-iqA_{\mu})(\partial^{\mu}-iqA^{\mu})+m^{2}]\phi(t,x)=0,\label{basicequation}
 \eea
where, $A_{\mu}=(0,0,0,A_{z}(t))$ and $\phi$ is the scalar field.

\begin{figure}[t]
    \center
    \includegraphics[width=0.7\columnwidth]{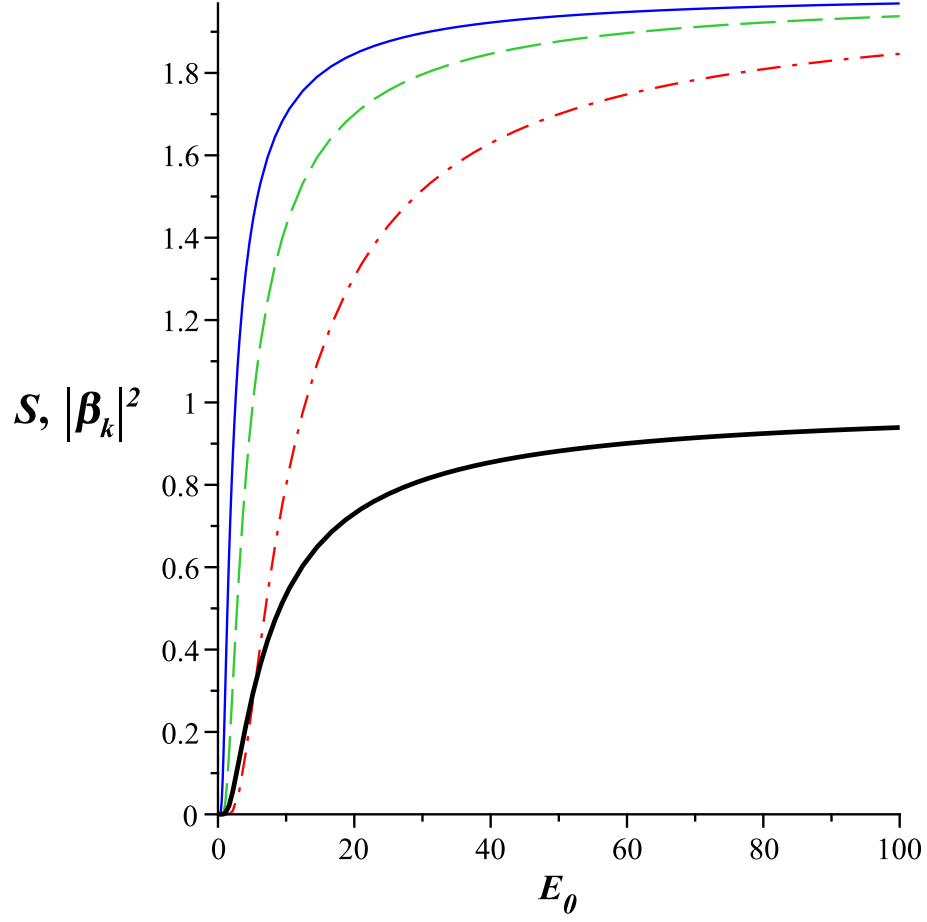}\\
    \caption{(Color online) The von Neumann entropy for bosonic modes with respect to the electric field for ($q,k_{\perp}=1$) and certain values of  mass: $m=0$ (solid line), $m=1$ (green dashed line), and $m=2$  (red dash-dotted line). $|\beta|^{2}$ as a function of $E_{0}$ is represented by the solid thick black line for $m=2$. In the large electric fields, $|\beta|^{2}$ tends to its maximum value, $|\beta_{k}|^{2}\rightarrow1$. Thus, regarding Eq. (\ref{sbetta}), entropy is a function of $|\beta_{k}|^{2}$ and at large values of the electric fields $E_{0}$ tends to a constant value $(S_{k}=2)$. }\label{figE}
   \end{figure}

 For the purpose of the present subsection, we restrict ourselves to the constant electric field and rewrite  Eq. (\ref{basicequation}) for $E_z(t)=E_{0}$:
\bea
[\partial^{2}_{t}+m^{2}+{\hat k}^{2}_{\perp}+({\hat k}_{z}-qE_{0}t)^{2}]\phi_{ks}(t,x)=0\label{basic}
 \eea
In the above equation,  natural units are used in which $c=\hbar=1$  and $A_{z}(t)$ is replaced by $A_{z}(t)=-E_{0}t$. This equation is used for scalar particles with zero spin. After turning to the momentum space, we have
\bea
\phi(t,r)=(2\pi)^{-3/2}\int dk \exp(ik.r)\widetilde\phi(t,k).
\eea
Therefore, Eq. (\ref{basic}) in the momentum space is given by
  \bea
  [\partial^{2}_{t}+m^{2}+k^{2}_{\perp}+(k_{z}-qE_{0}t)^{2}]\widetilde\phi_{k}(t)=0\label{equation}
  \eea
where, $\widetilde\phi_{k}(t)=\widetilde\phi(t,k)$ is the Fourier component of the Klein-Gordon equation for scalar particles and
   $k^{2}_{\perp}=k^{2}_{x}+k^{2}_{y}$. Changing to the following convenient variables
\bea
z&=&\sqrt{2}\xi e^{i\pi/4},\   \ \xi=\frac{(k_{z}-qE_{0}t)}{\sqrt{qE_0}},\nonumber\\
\nu&=&-\frac{1}{2}-i\frac{\mu}{2},\   \ \mu=\frac{m^{2}+k^{2}_{\perp}}{qE_0},\label{zp}
\eea
Eq. (\ref{equation}) will be  converted to the following equation:
\bea
[\partial^{2}_{z}+(\nu+\frac{1}{2}-\frac{z^{2}}{4})]\widetilde\phi_{\nu}(z)=0\label{solution}.
 \eea
  The solutions of Eq. (\ref{solution}) are the parabolic cylinder functions denoted by the symbol $D_{\nu}(z)$ \cite{gradshteyn}
 \bea
 D_{\nu}(z)=2^{\frac{\nu-1}{2}}e^{\frac{-z^{2}}{4}}\Psi(\frac{1-\nu}{2},\frac{3}{2},\frac{z^{2}}{2})\label{confluent}
 \eea
  where, $\Psi (a,b,z)$ is the confluent hypergeometric function. The functions $D_{\nu}(-z)$ and $D_{-\nu-1}(\pm iz)$ also satisfy Eq. (\ref{solution})
  \cite{gradshteyn}.  The following linear relations between parabolic cylinder functions show how any three of the solutions are connected:
    \bea
D_{\nu}(z)&=&\frac{\Gamma(\nu+1)}{2\pi}[e^{\pi/2}D_{-\nu-1}(iz)+e^{\frac{-i\pi \nu}{2}}D_{-\nu-1}(-iz)]\nonumber\\
&=&\frac{\sqrt{2\pi}}{\Gamma(-\nu)}e^{\frac{-i\pi (\nu+1)}{2}}D_{-\nu-1}(iz)+e^{-i\pi \nu}D_{\nu}(-z).\label{linear}
\eea
     Therefore, there are precisely two linearly independent solutions of Eq. (\ref{solution}). For all values of $\nu$, $D_{\nu}(z)$ and $D_{-\nu-1}(\pm iz)$ are linearly independent. In order to calculate entanglement, we need the asymptotic solutions at $t_{in}\rightarrow - \infty$ and $t_{out}\rightarrow + \infty$  because we are interested in solutions with negative and positive frequencies. The asymptotic behavior of the solutions for large values of $|z|$ is given by \cite{gradshteyn}
     \bea
      D_{\nu}(z)&\approx& e^{-z^{2}/4}z^{\nu},(\mid z\mid\gg\mid \nu\mid, \mid arg (z)\mid<\frac{3\pi}{4})
      \eea
 Using Eqs. (\ref{solution}) and (\ref{confluent}), one can find the asymptotic solution at $t_{in}=-\infty$ for a particle with momentum $k$ and charge $q$
 \bea
D_{\nu}(z)&=&D_{-\frac{1}{2}-i\frac{\mu}{2}}({\sqrt{2}}\xi e^{i\pi/4})\nonumber\\
&\approx&(2\xi^{2})^{(-i\mu-1)/4}e^{(\mu-i)\pi/8}e^{-i\xi^{2}/2}\label{particle},
 \eea
 where, $\xi\gg1$.
  Using $k\rightarrow -k$ and $q\rightarrow -q$ in Eq. (\ref{equation}), we obtain another solution with negative frequency which describes an incoming antiparticle as below:
 \bea
 D_{-\nu-1}(-iz)=D_{-\frac{1}{2}+i\frac{\mu}{2}}(\sqrt{2}\xi e^{-i\pi/4})\nonumber\\
 \approx(2\xi^{2})^{(i\mu-1)/4}e^{(\mu+i)\pi/8}e^{i\xi^{2}/2}\label{antiparticle}.
 \eea
  In these solutions, the asymptotic phases and frequencies are:
 \bea
\pm\frac{1}{2}\xi^{2}&=& \frac{1}{2qE_0}k_{z}^2-k_{z}t+\frac{1}{2}qE_0t^2,\nonumber\\
 \pm\partial_{t}\frac{1}{2} {\xi^{2}}&=&\pm(-k_{z}+qE_0t)\sim\pm \omega.
 \eea
 We can also find the sets of solutions at $t_{out}=+\infty$. For an outgoing particle with momentum $k$ and charge $q$, the convenient solution is
 \bea
 D_{-\nu-1}(iz)=D_{-\frac{1}{2}-i\frac{\mu}{2}}(\sqrt{2}|\xi|e^{-i\pi/4}),
 \eea
 where, $|\xi|=\frac{-k_{z}+qE_0t}{\sqrt{qE_0}}\gg1$. In the same manner, $D_{\nu}(-z)$ describes an outgoing antiparticle.
 Using Bogoliubov transformation \cite{carol}, one can expand the sets of solutions at $t_{in}=- \infty$ in terms of the sets of solutions at $t_{out}=+ \infty$ as follows:
\bea
\phi_{in,k}^{+}=\alpha_{k} \phi_{out,k}^{+}+\beta_{k}\phi_{out,k}^{-}\label{inout},
\eea
where, $\alpha_{k}$ and $\beta_{k}$ are Bogoliubov coefficients. The '$in$' positive frequency mode $\phi_{in,k}^{+}=D_{\nu,in}(z)$ is expressed as a linear combination of the '$out$' positive $\phi_{out,k}^{+}=D_{-\nu-1,out}(iz)$ and negative $\phi_{out,k}^{-}=D_{\nu,out}(-z)$ frequency modes.    Using a linear relation between $D_{\nu}(\pm z)$ and $D_{-\nu-1}(\pm iz)$, Eq. (\ref{linear}), one can achieve the Bogoliubov coefficients as follows:
 \bea
 \alpha_{k}=\frac{\sqrt{2\pi}}{\Gamma(-\nu)}e^{\frac{-i\pi (\nu+1)}{2}},\    \ \beta_{k}=e^{-i\pi \nu}.\label{alpha}
 \eea
 Taking into account\cite{gradshteyn}
 \bea
 \frac{\pi}{|\Gamma(\frac{1}{2}+ix)|^{2}}=\cosh(\pi x),
 \eea
 these coefficients for scalar particles will satisfy the following relation
\bea
|\alpha_{k}|^{2}-|\beta_{k}|^{2}=1.\label{bs}
\eea
Now, we calculate the entanglement which is generated by the background constant electric field.
 It is necessary to specify the '$in$' and '$out$' states and operators. The operators $a_{k,in},b_{k,in}$ and $a_{k,out},b_{k,out}$
annihilate the '$in$' $|0_{k}0_{-k}\rangle_{in}$ and '$out$'  $|0_{k}0_{-k}\rangle_{out}$ vacuum for each momentum, respectively.
\bea
a_{k,in}|0_{k}0_{-k}\rangle_{in}&=&b_{k,in}|0_{k}0_{-k}\rangle_{in}=0\nonumber\\
a_{k,out}|0_{k}0_{-k}\rangle_{out}&=&b_{k,out}|0_{k}0_{-k}\rangle_{out}=0\label{vacu}
\eea
where, the (k,-k) subscripts indicate the particle and antiparticle modes. Using the Bogoliubov transformation, the relation
between these operators is given by \cite{carol}
\bea
a_{k,in}&=&\alpha^{*}_{k}\ \ a_{k,out}-\beta^{*}_{k}\ \ b^{\dagger}_{k,out}\nonumber\\
b_{k,in}&=&\alpha^{*}_{k}\ \ b_{k,out}-\beta^{*}_{k}\ \ a^{\dagger}_{k,out}\label{oprator}
\eea
Now, using the convenient calculations, we show that the separable '$in$' states can be expanded in terms of the '$out$' entangled state.
The state vector of the system can be described by the tensor product of
the two Hilbert spaces $H_{k}\bigotimes H_{k'}$, where $H_{k}$ indicates the Hilbert space related to particles and $H_{k'}$ to the antiparticles  created by the electric field.
 The $in$-vacuum state is defined by the absence of any mode excitations
$$|0\rangle_{in}=\prod_{kk'} (|0_{k}\rangle|0_{k'}\rangle)_{in},$$
Using the Schmidt decomposition, the in-vacuum state for each mode can be expanded in terms of the out-states \cite{Schmidt}
 $$(|0_{k}\rangle|0_{-k}\rangle)_{in}=\sum_{n}c_{n}(|n_{k}\rangle|n_{-k}\rangle)_{out},$$
where, $n$ indicates the number of particles with momentum k and the number of antiparticles with momentum $-k$   created by the electric field.
For simplicity, $|n_{k}\rangle|n_{-k}\rangle$ is replaced by
 $|n_{k} n_{-k}\rangle$, and we will, therefore, have
  \bea
 |0_{k}0_{-k}\rangle_{in}=\sum_{n}c_{n}|n_{k}n_{-k}\rangle_{out}.\label{schmit}
 \eea
The $|0_{k}0_{-k}\rangle_{in}$ is a separable state from the  view point of an observer in the $in$-region.
  If there are  more than one non-zero coefficients on the right hand side of Eq.(\ref{schmit}),   then the separable $in$-state is the entangled state from the view point of  an inertial observer in the $out$-region. Therefore, we have to determine $c_{n}$ to evaluate the measure of the entanglement. For this purpose, we use the definition of vacuum and its normalization.
      Substituting Eqs. (\ref{oprator}) and (\ref{schmit}) in the definition of vacuum
   \bea
  & a_{k,in}&|0_{k}0_{-k}\rangle_{in}=0\nonumber\\
   (\alpha^{*}_{k}\ \ a_{k,out}-\beta^{*}_{k} &b^{\dagger}_{k,out}&)\sum_{n}c_{n}|n_{k}n_{-k}\rangle_{out}=0,\label{difi}
  \eea
 leads to
\bea
 c_{n+1}=\frac{\beta^{*}_{k}}{\alpha^{*}_{k}}c_{n},
\eea
Normalization of vacuum, $$\langle 0_{k}0_{-k}|0_{k}0_{-k}\rangle_{in}=1$$ leads to
 \bea
\sum_{n} |c_{n}|^{2}&=&1\nonumber\\
&=&|c_{0}|^{2}(1+|\frac{\beta_{k}}{\alpha_{k}}|^{2}+|\frac{\beta_{k}}{\alpha_{k}}|^{4}+...).\label{vaccum}
\eea
Thus, the coefficients $c_{n}$ are given by:
\bea
|c_{0}|^{2}&=&|\frac{1}{\alpha_{k}}|^{2}\nonumber\\
|c_{n}|^{2}&=&|\frac{\beta_{k}}{\alpha_{k}}|^{2n}|c_{0}|^{2}\nonumber\\
&=&(1-|c_{0}|^{2})^{n}|c_{0}|^{2}\label{cn}
\eea
Based on  the values of $c_{n}$ thus obtained, we expect an entanglement generation to occur. We can utilize an appropriate measure of  entanglement, namely  the
von Neumann entropy  defined as follows:
\bea
S(\rho_{k})=-Tr(\rho_{k}\log_{2}(\rho_{k})).\label{entropy}
\eea
First, we have to specify the density matrix of the whole system, $\rho_{k,-k}$,  followed by  reduced density matrix of the subsystem, $\rho_{k}$. All the
 properties of the system can be deduced from the density matrix
 \bea
 \rho_{k,-k}&=&|0_{k}0_{-k}\rangle_{in}\langle 0_{k}0_{-k}|\nonumber\\
 &=&\sum_{n,m}c_{n}c^{*}_{m}|n_{k}n_{-k}\rangle_{out}\langle m_{k}m_{-k}|.
 \eea
 As we wish to deal with only one of the subsystems, we use the concept of  reduced density matrix. One can find the reduced density matrix for the
 subsystem related to the particles (denoted by k), obtained by tracing $\rho_{k,-k}$ over all the states of the subsystem related to the antiparticles
 (denoted by -k), so that
 \bea
 \rho_{k}&=&Tr_{-k}(\rho_{k,-k})\nonumber\\
 &=&\sum_{l}\langle l_{-k}|\rho_{k,-k}|l_{-k}\rangle\nonumber\\
 &=&\sum_{n}|c_{n}|^{2}|n_{k}\rangle\langle n_{k}|.\label{reducedmatrix}
 \eea
 The von Neumann entropy for scalar modes described by Eq. (\ref{entropy}) is given by
 \bea
 S_{k}&=&-\sum_{n}|c_{n}|^{2}\log_{2}|c_{n}|^{2}\nonumber\\
 &=&\log_{2}\frac{x^{\frac{x}{x-1}}}{1-x}\label{scalarentropy},
 \eea
where, $x=|\frac{\beta_{k}}{\alpha_{k}}|^{2}$ and is determined by Eq. (\ref{alpha})
\bea
x=\frac{|\beta_{k}|^{2}}{1+|\beta_{k}|^{2}},\      \ |\beta_{k}|^{2}= e^{\frac{-\pi(m^{2}+k_{\perp}^{2})}{qE_{0}}}.\label{betaalpha}
\eea
 Therefore, we get the von Neumann entropy with respect to the electric field, transverse components
 of momentum, as well as particle's mass and charge. Eq. (\ref{scalarentropy}) can be written in terms of $|\beta_{k}|^{2}$
 \bea
  S_{k}=-|\beta_{k}|^{2} \log_{2}|\beta_{k}|^{2}+(1+|\beta_{k}|^{2})\log_{2}(1+|\beta_{k}|^{2})\label{sbetta}.
  \eea
  According to Eq. (\ref{sbetta}), the increase in $|\beta_{k}|^{2}$ value enhances the von Neumann entropy. Both the von Neumann entropy, $S_{k}$, and $|\beta_{k}|^{2}$ are increasing functions with respect to $E_{0}$. The variation of the $S_{k}$, and $|\beta_{k}|^{2}$ as a function of the electric field $E_{0}$ is shown in Fig. \ref{figE}. In the large electric fields, $|\beta_{k}|^{2}$ tends to its maximum value, $|\beta_{k}|^{2}\rightarrow1$. Thus, regarding Eq. (\ref{sbetta}), entropy is a function of $|\beta_{k}|^{2}$ and at large values of the electric fields $E_{0}$ tends to a constant value $(S_{k}=2)$.

\begin{figure}[b]
    \center
    \includegraphics[width=0.7\columnwidth]{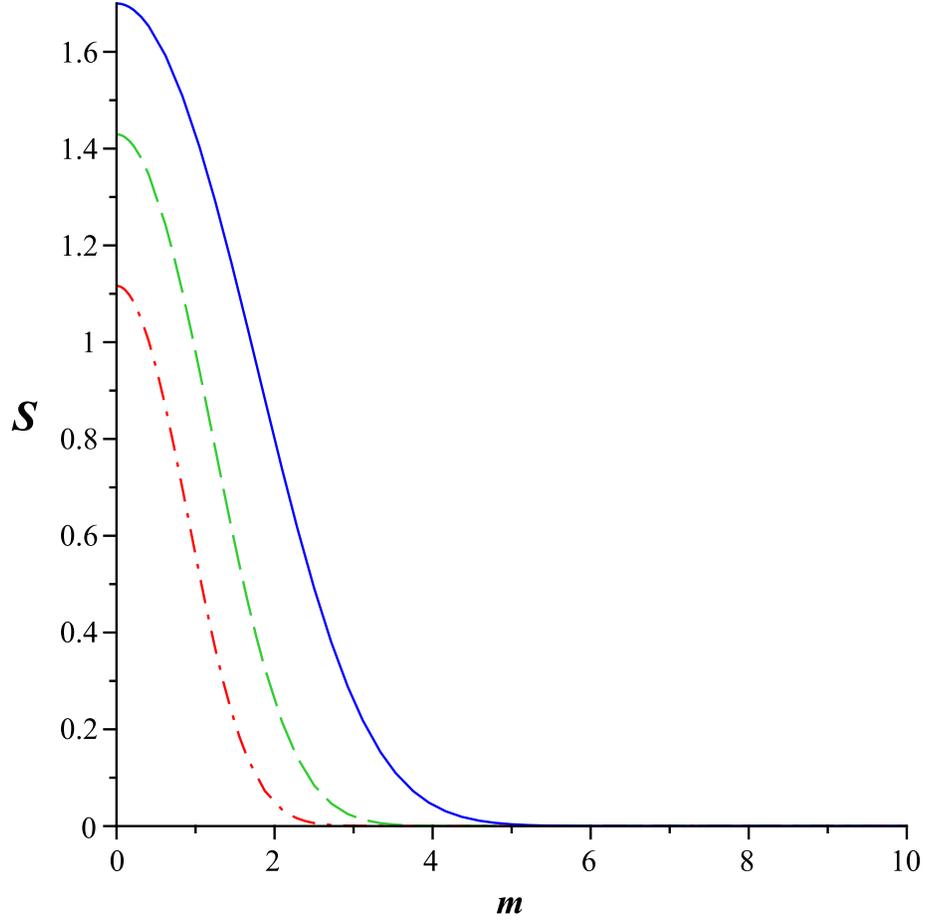}\\
    \caption{(Color online) The von Neumann entropy for bosonic modes with respect to particle  mass for ($q,k_{\perp}=1$) and certain values of the electric field:  $E=10$ (solid line), $E=5$  (green dashed line), and $E=3$  (red dash-dotted line). The maximum value of entanglement for fixed values of $q$, $k_{\perp}$ and $E_{0}$ occurs in $m=0$ which is obtained by substituting $|\beta_{k}|^{2}=\exp(\frac{-\pi k_{\perp}^{2}}{2qE_{0}})$ in Eq. (\ref{sbetta}). }\label{figm}
   \end{figure}

  $|\beta_{k}|^2$ is the mean number of the particles (antiparticles) produced in mode $k$ $(-k)$
 \bea
 \langle 0_{k}0_{-k}|a_{out,k}^{\dagger} a_{out,k}|0_{k}0_{-k}\rangle_{in}=|\beta_{k}|^2.\label{production}
 \eea
   When the mean number of the produced pairs increases, the entanglement generated for bosonic modes will also increase. $S_{k}$ and $|\beta_{k}|^{2}$ exhibit similar behaviors for bosonic modes.
 The variation of entanglement with respect to mass is shown in Fig. \ref{figm}. A specific electric field creates more particles of smaller mass than those of larger mass. Fig. \ref{figm} indicates that the measure of entanglement for fixed values of $q$, $k_{\perp}$ and $E_{0}$ is greater for particles with smaller mass than it is for those of larger mass. The maximum value of entanglement for fixed values of $q$, $k_{\perp}$ and $E_{0}$ occurs in $m=0$ which is obtained by substituting $|\beta_{k}|^{2}=\exp(\frac{-\pi k_{\perp}^{2}}{2qE_{0}})$ in Eq. (\ref{sbetta}).
 Since $m$ and $k_\perp$ appear in the same form in $S_{k}$, the entanglement behavior will be similar  with respect to $k_{\perp}$ and $m$.

In Eq. (\ref{schmit}), we express $in$-vacuum in terms of $out$-states. The probability of vacuum-to-vacuum transition is given by  
\bea
|\langle 0,out|0,in\rangle|^{2}=|c_{0}|^{2}=\frac{1}{|\alpha_{k}|^{2}}.
\eea
The maximum value of $|c_{0}|^{2}=1$; this means that the vacuums of the '$in$' and '$out$' regions are the same. Decreasing  value of $|c_{0}|^{2}$ means that the initial vacuum decays to the more pairs in the '$out$' region. Therefore, it is reasonable to suggest that a smaller value of $|c_{0}^{2}|$ leads to a more entangled state.
Since the value of $|\beta|^{2}$ ranges between $0$ and $1$, and also because $|\alpha_{k}^{2}|=1+|\beta|^{2}$, the minimum value of $|c_{0}^{2}|$ occurs at $|\beta|^{2}=1$. 

\subsection{ Entanglement Entropy for Fermion Particles }
 In this subsection, we will investigate  the generation of  entanglement for fermionic modes. We use asymptotic solution of equation of motion for charged fermion particles in the presence of an electric field.
  Consider an electric field along the $z$-direction. It is related to the gauge potential through $E_{z}(t)=-\partial A_{z}(t)/\partial t$. For a particle of mass $m$ and charge $q$, the Dirac equation on the four dimensional Minkowski spacetime with the metric (+,-,-,-) is given by
\bea
 (i\gamma^{\mu}\partial_{\mu}-q\gamma^{\mu} A_{\mu}-m)\Psi(x)=0\label{dirac}
 \eea
where, $A_{\mu}=(0,0,0,A_{z}(t))$. $\gamma^{\mu}$ and $\Psi$ are the Dirac matrix and spinors, respectively \cite{gama}.
One can introduce $\Psi(x)$ to have:
\bea
\Psi(x)=(i\gamma^{\nu}\partial_{\nu}-q\gamma^{\nu} A_{\nu}+m)R(x)\label{psi}
\eea
The second order differential equation is
\bea
&(&i\gamma^{\mu}\partial_{\mu}-q\gamma^{\mu} A_{\mu}-m)(i\gamma^{\nu}\partial_{\nu}-q\gamma^{\nu} A_{\nu}+m)R \label{equ}\nonumber\\
=&[&-\partial^{2}_{t}-m^{2}-\partial^{2}_{x}-\partial^{2}_{y}-(\partial_{z}-iqA_{z})^{2}+iqE_{z}\alpha^{3}] R\nonumber\\
\eea
 where, $\alpha^{3}=\gamma^{0}\gamma^{3}=\left(
             \begin{array}{cc}
               0 & \sigma_{3} \\
                \sigma_{3} &0 \\
             \end{array}
           \right)$.
           We search for a solution of Eq. (\ref{equ}) of the following form:
 \bea
 R(x)&=&\varphi(t,\mathbf{x}) \chi\nonumber\\
 \varphi(t,\mathbf{x})&=&(2\pi)^{-3/2}\int e^{i\mathbf{k}.\mathbf{x}} \phi (t,\mathbf{k}) d\mathbf{k}\label{requ}
 \eea
where, $\varphi(x)$ is a complex scalar function and $\chi$ designates the  eigenbispinors of the  $\alpha^{3}$ and $\Sigma_{3}=\frac{i}{2}\gamma^1\gamma^2$ :  
 \bea
\chi_{+}^{\uparrow}&=&\left(
   \begin{array}{c}
     0\\
     1\\
     0\\
    -1\\
   \end{array}
 \right),\chi_{+}^{\downarrow}=\left(
   \begin{array}{c}
     1\\
     0\\
     1\\
     0\\
   \end{array}
 \right),\nonumber\\
 \chi_{-}^{\uparrow}&=&\left(
   \begin{array}{c}
     1\\
     0\\
    -1\\
     0\\
   \end{array}
 \right),\chi_{-}^{\downarrow}=\left(
   \begin{array}{c}
     0\\
     1\\
     0\\
     1\\
   \end{array}
 \right),\nonumber\\
 &&\alpha^{3}\chi_{\pm}^{s}=\chi_{\pm}^{s},~~ (s\in\{\uparrow,\downarrow\})\nonumber\\
&&{\Sigma_{3}}\chi_{\pm}^{\uparrow}=+\chi_{\pm}^{\uparrow},~~\Sigma_{3}\chi_{\pm}^{\downarrow}=-\chi_{\pm}^{\downarrow},
\label{eigen}
 \eea
$\Sigma_{3}$  is the matrix of the spin component along the direction of the electric field and commutes with $\alpha^{3}$.

 Using Eqs. (\ref{equ}-\ref{eigen}) and substituting the standard representation for the Dirac matrix, we have
\bea
 [\partial^{2}_{t}+m^{2}+k^{2}_{\perp}+(k_{z}+qA_{z}(t))^{2}+ iqE(t)] \phi_{k,s}(t)\chi_{+}^{s}=0\label{second}\nonumber\\
 \eea
where, $\phi_{k,s}(t)\chi_{+}^{s}$ and $\phi_{k,s}^{*}(t)\chi_{+}^{s}$ specify the spin-up and down particle and antiparticle, respectively.
Then, the solutions of Eq. (\ref{second}) form a complete set.
Another solution of the second order differential equation (\ref{equ}) with negative eigenvalue , $- iqE(t)$, satisfies the following equation
 \bea
  [\partial^{2}_{t}+m^{2}+k^{2}_{\perp}+(k_{z}+qA_{z}(t))^{2}- iqE(t)] \phi_{k,s}(t)\chi_{-}^{s}=0\label{secondorder}\nonumber\\
 \eea
The second order differential equation (\ref{equ}) leads to Eqs.(\ref{second}) and (\ref{secondorder}), while the Dirac equation is a first order one. Therefore, it will suffice to have one complete set of solutions corresponding to either (\ref{second}) or (\ref{secondorder}). In fact, Eq. (\ref{secondorder}) does not lead to any new result.  Therefore, we consider Eq. (\ref{second}) and write $R$ in the following form
\bea
R(t,k)&=&\sum_{s} (c_{1}^{s}\phi_{k,s}(t)\chi_{+}^{s}+c_{2}^{s}\phi_{k,s}^{*}(t)\chi_{+}^{s})\label{R}
\eea
 Using (\ref{psi}), $\Psi$ takes the following form
 \bea
\Psi(t,k)=\sum_{k,s}(a_{k,s}u_{k}^{s}(x)+ b^{\dagger}_{k,s} v_{k}^{s}(x))
\eea
with
\bea
u_{k}^{s}(t)&=&(i\gamma^{0}\partial_{t}-\vec{\gamma}.(\vec{k}+q\vec{A})+m)\phi(t,k)  \chi_{+}^{s}\nonumber\\
v_{k}^{s}(t)&=&(i\gamma^{0}\partial_{t}-\vec{\gamma}.(\vec{k}+q\vec{A})+m)\phi(t,k)^{*}  \chi_{+}^{s}\label{uv}
\eea
  According to Eq. (\ref{solution}), $ \phi_{k,s}(t)$ in Eq. (\ref{second}) are
   parabolic cylinder functions
   \bea
    D_{\nu}(\pm z),  D_{-\nu-1}(\pm iz) \nonumber\\
    \nu=-1-i\frac{\mu}{2} ,\     \mu=\frac{m^2+k^{2}_{\perp}}{q E}.\label{para}
    \eea
Using Eqs. (\ref{uv}) and (\ref{para}) and the invariant inner product 
\bea
(f_{k}^{r}(t,x),g_{p}^{s}(t,x))=\int (f_{k}^{r}(t,x))^\dagger g_{p}^{s}(t,x)d^3x\label{innerproduct}
 \eea
  one can evaluate the Bogoliubov coefficients for Dirac's fermions in a background constant electric
 field as follows
\bea
(u^{s,in}_{k},u^{r,out}_{p})=\delta_{rs}\delta(\vec{k}-\vec{p})\alpha^{s}_{k}\nonumber\\
(u^{s,in}_{k},v^{r,out}_{p})=\delta_{rs}\delta(\vec{k}-\vec{p})\beta^{s}_{k} ,\label{fabb}
\eea
with
\bea
\alpha^{s}_{k}&=&\alpha^{\uparrow}_{k}=\alpha^{\downarrow}_{k}=\sqrt{\frac{\mu}{\pi}}\Gamma(\frac{i\mu}{2}) \sinh(\frac{\pi\mu}{2})e^{-\frac{\pi\mu}{4}}\nonumber\\\label{fab}
\beta^{s}_{k}&=&\beta^{\uparrow}_{k}=\beta^{\downarrow}_{k}=e^{-\pi\frac{\mu}{2}}.
\eea
As indicated in Eq. (\ref{eigen}) , $s\in\{\uparrow,\downarrow\}$ is related to the positive and negative eigenvalues of the matrix of the spin component along the direction of the electric field. Since the spin has no interaction with the electric field,  the Bogoliubov coefficients for the up and down spins are the same.

Taking \cite{gradshteyn} into account
\bea
|\Gamma(ix)|^{2}=\frac{\pi}{x\sinh \pi x}
\eea
these coefficients will satisfy the relation below
\bea
|\alpha|^{2}+|\beta|^{2}=1\label{relation}
\eea
  The relationship between the '$in$' and the '$out$' operators is expressed by
 \bea
 a_{d,out}=\alpha_{d}\ \ a_{d,in}-\beta^{*}_{d}\ \ b^{\dagger}_{d,in}\nonumber\\
 b^{\dagger}_{d,out}=\alpha^{*}_{d}\ \ b^{\dagger}_{d,in}+\beta_{d}\ \ a_{d,in},\label{abspinor}
 \eea

 \noindent where, $a_{d}$ and $b_{d}$ are the annihilation operators for particle and antiparticle, respectively, and subscript $d$ stands for momentum $k$ and
  spin $s\in\{\uparrow,\downarrow\}$.
 The vacuum state is given by
\bea
|0\rangle_{in}=\prod_{k,k',s}(|0^{s}_{k}\rangle|0^{s}_{k'}\rangle)_{in}.
\eea
 Using the Schmidt decomposition and Pauli exclusion principle, we can expand the '$in$' vacuum state in terms of the '$out$' states for a single mode $k$
 \bea
|0_{k}0_{-k}\rangle_{in}&\equiv&(|0_{k}^{\uparrow}0_{-k}^{\downarrow}\rangle|0_{k}^{\downarrow}0_{-k}^{\uparrow}\rangle)_{in}=\sum_{n=0,1}c'_{n}|n_{k}^{\uparrow}n_{-k}^{\downarrow}\rangle_{out}
\sum_{m=0,1}c^{''}_{m}|m_{k}^{\downarrow}m_{-k}^{\uparrow}\rangle_{out} \nonumber\\
&=&c_{0}|0^{\uparrow}_{k}0^{\downarrow}_{-k}\rangle_{out}|0^{\downarrow}_{k}0^{\uparrow}_{-k}\rangle_{out}
 + c_{1}|1^{\uparrow}_{k}1^{\downarrow}_{-k}\rangle_{out}|1^{\downarrow}_{k}1^{\uparrow}_{-k}\rangle_{out}\nonumber\\
 &+&
 c_{2} |1^{\uparrow}_{k}1^{\downarrow}_{-k}\rangle_{out}|0^{\downarrow}_{k}0^{\uparrow}_{-k}\rangle_{out}
 +
 c_{3} |0^{\uparrow}_{k}0^{\downarrow}_{-k}\rangle_{out} |1^{\downarrow}_{k}1^{\uparrow}_{-k}\rangle_{out} ,\label{fermionstate}
 \eea
 where, the coefficients $c'_{n}c^{''}_{m}$ are replaced to $c_{i}$ and symbol $\uparrow$ $(\downarrow)$ indicate up(down) spin.
 Imposing  $a_{d,in}|0_{k}0_{-k}\rangle_{in}=0$, $(\langle 0_{k}0_{-k}|0_{k}0_{-k}\rangle)_{in}=1$ and using Eq. (\ref{abspinor}),
we obtain the four coefficients $c_{i}$
\bea
|c_{0}|^{2}&=&|\alpha^{\uparrow}_{k}|^{2}|\alpha^{\downarrow}_{k}|^{2},\ \ |c_{1}|^{2}=|\beta^{\uparrow}_{k}|^{2}|\beta^{\downarrow}_{k}|^{2}\nonumber\\
|c_{2}|^{2}&=&|\alpha^{\uparrow}_{k}|^{2}|\beta^{\downarrow}_{k}|^{2},\ \ |c_{3}|^{2}=|\alpha^{\downarrow}_{k}|^{2}|\beta^{\uparrow}_{k}|^{2}.\label{cnf}
\eea
The states in (\ref{fermionstate}) are designated by A, B, C and D.
\bea
 |0_{A}\rangle_{out}&\equiv&|0^{\uparrow}_{k}\rangle_{out},\  \ |1_{A}\rangle_{out}\equiv|1^{\uparrow}_{k}\rangle_{out},\nonumber\\
 |0_{B}\rangle_{out}&\equiv&|0^{\downarrow}_{k}\rangle_{out},\  \ |1^{\downarrow}_{k}\rangle_{out}\equiv|1_{B}\rangle_{out},\nonumber\\
|0_{C}\rangle_{out}&\equiv&|0^{\uparrow}_{-k}\rangle_{out},\  \ |1^{\uparrow}_{-k}\rangle_{out}\equiv|1_{C}\rangle_{out},\nonumber\\
|0_{D}\rangle_{out}&\equiv& |0^{\downarrow}_{-k}\rangle_{out},\  \ |1^{\downarrow}_{-k}\rangle_{out}\equiv|1_{D}\rangle_{out},\nonumber\\\label{ABCD}
\eea
Using the representation  (\ref{ABCD}), the '$in$' vacuum state is expressed by
\bea
(|0_{k}^{\uparrow}0_{-k}^{\downarrow}\rangle|0_{k}^{\downarrow}0_{-k}^{\uparrow}\rangle)_{in}=|\Psi_{ABCD}\rangle_{out}.
\eea
We can calculate the measure of entanglement between the one part of system to the rest.

 $S_{A(BCD)}$, for example, is the measure of entanglement between the state A and the states B, C, D as follows
\bea
S_{A(BCD)}=-Tr(\rho_{A}\log_{2}\rho_{A}).\label{sa}
\eea
The reduced density operator $\rho_{A}$ in $S_{A(BCD)}$ after tracing on B, C, and D is given by
\bea
\rho_{A}&=&Tr_{(BCD)}(|\Psi_{ABCD}\rangle_{out}\langle\Psi_{ABCD}|)\nonumber\\
&=&(|c_{0}|^{2}+|c_{3}|^{2})|0_{A}\rangle_{out}\langle 0_{A}|\nonumber\\
&+&(|c_{1}|^{2}+|c_{2}|^{2})|1_{A}\rangle_{out}\langle 1_{A}|\nonumber\\
&=&|\alpha_{k}|^{2}|0_{A}\rangle_{out}\langle 0_{A}|+|\beta_{k}|^{2}|1_{A}\rangle_{out}\langle 1_{A}|.\label{ro}
\eea
$S_{A(BCD)}$, the entanglement entropy between a spin up particle with mode $k$ to the rest of the system is obtained by

\bea
S_{A(BCD)}=-|\alpha_{k}|^{2}\log_{2}(|\alpha_{k}|^{2})-|\beta_{k}|^{2}\log_{2}(|\beta_{k}|^{2}).
\eea

In the same manner, $S_{B(ACD)}$, $S_{C(ABD)}$ and $S_{D(ABC)}$ can be calculated and their values are equal to $S_{A(BCD)}$
\bea
S_{k}=-|\alpha_{k}|^{2}\log_{2}(|\alpha_{k}|^{2})-|\beta_{k}|^{2}\log_{2}(|\beta_{k}|^{2}),\    \ |\beta_{k}|^{2}=e^{\frac{-\pi (m^{2}+k_{\perp}^{2})}{2qE_{0}}},\  \ |\alpha_{k}|^{2}=1+|\beta_{k}|^{2}\label{ssfermion}
 \eea
  We can also get the average von Neumann entropy \cite{gilad} as follows
\bea
S=\frac{1}{4}(S_{A(BCD)}+S_{B(ACD)}+S_{C(ABD)}+S_{D(ABC)})\label{sfermion}
\eea
  As expected, all of the entanglement entropies in Eq. (\ref{sfermion}) have the same value, because the electric field can not distinguish spin up and down particles. In other words, each of the von Neumann entropies in Eq. (\ref{sfermion}) means the entanglement between one part of the system which is a particle(antiparticle) with mode $k$ and spin $s$ with the rest of the system.

\begin{figure}[t]
    \center
    \includegraphics[width=0.7\columnwidth]{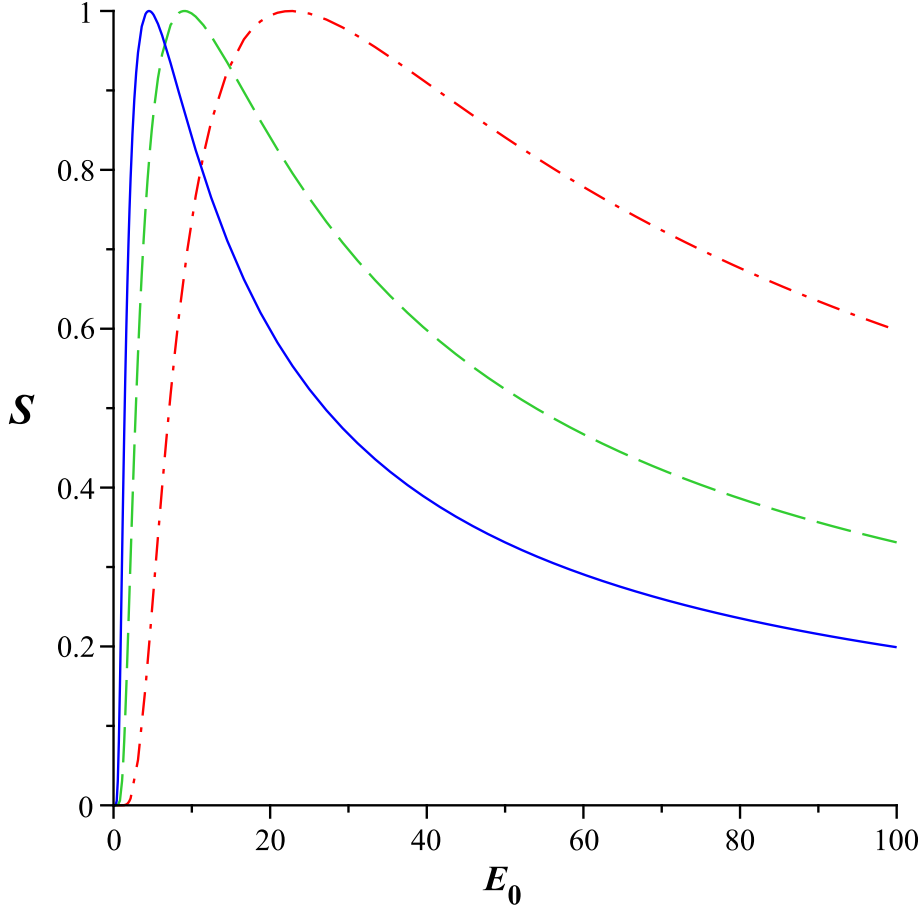}\\
    \caption{(Color online) The von Neumann entropy for fermionic modes with respect to the electric field for ($q,k_{\perp}=1$) and certain values of the fermion mass: $m=0$ (solid line),  $m=1$ (green dashed line), and  $m=2$ (red dash-dotted line). For very small or large electric fields, the value of $|\beta_{k}|^{2}$ tends to $0$ or $1$, respectively. Therefore, according to Eq. (\ref{ssfermion}) the entropy has its minimum value, $S_{min}=0$. The maximum value of entropy, $S_{max}=1$, for fixed values of $q$, $k_{\perp}$ and $m$,
     occurs in$ E_{0}=\frac {\pi (m^{2}+k_{\perp}^{2})}{q\ln(2)}$, which is equivalent to $|\beta_{k}|^{2}=\frac{1}{2}$.
   }\label{figEf}
   \end{figure}

\begin{figure}[t]
    \center
    \includegraphics[width=0.7\columnwidth]{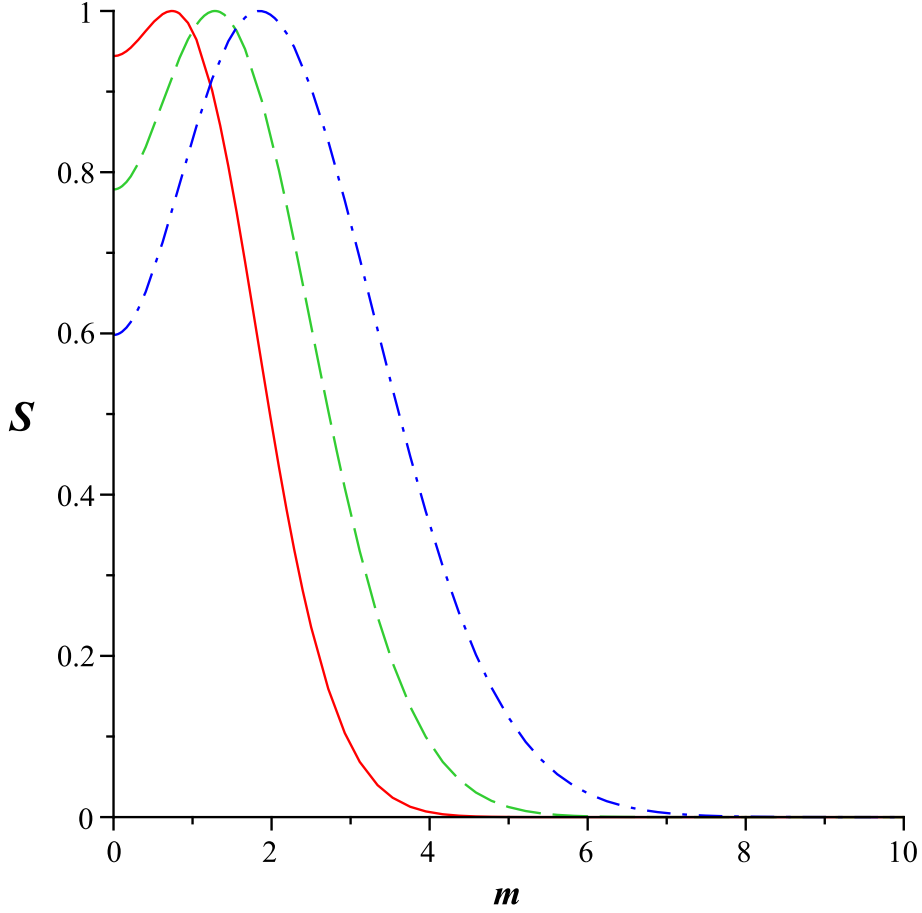}\\
    \caption{(Color online) The von Neumann entropy for fermionic modes with respect to the particle mass  for ($q, k_{\perp}=1$) and certain values of the electric field:  $E=7$ ( red solid line), $E=12$  (green dashed line), and  $E=20$ (blue dash-dotted line). The entanglement entropy is obtained by substituting $|\beta_{k}|^{2}=\exp(\frac{-\pi k_{\perp}^{2}}{2qE_{0}})$ in Eq. (\ref{ssfermion}), for massless particles. Large values of the electric fields correspond to large values of $|\beta_{k}|^2$ and therefore the small value of $S_{k}$. According to Eq. (\ref{ssfermion}), for fixed values of $E_{0}$, $k_{\perp}$ and $q$, the maximum value of entropy is equal to one for $m^{2}=\frac{E_{0}q\ln(2)}{\pi}-k_{\perp}^{2}$. }\label{figmf}
   \end{figure}

 Eqs. (\ref{fermionstate}) and (\ref{cnf}) indicate the
   expanding of $in$- vacuum in terms of '$out$' states. Based on (\ref{relation}), $|\alpha_{k}|^2$ and $|\beta_{k}|^2$ in fermionic modes range  between $0$ and $1$. If $|\alpha_{k}|^2$ and $|\beta_{k}|^2$ have a value of either zero or one, we will have a separable state that leads to zero entanglement. The maximum  entanglement occurs when all the coefficients $c_{n}$ in Eq. (\ref{fermionstate}) are equal and nonzero. Therefore, the entanglement will have its maximum value in $|\alpha_{k}|^2=|\beta_{k}|^2=\frac{1}{2}$.
The behavior of entanglement entropy for fermionic modes is shown in Figs. \ref{figEf} and \ref{figmf}. In very small or large electric fields the value of $|\beta_{k}|^2$ tends to $0$ or $(1)$, respectively; therefore, the entropy has its minimum value, $S_{min}=0$, as indicated in Fig. \ref{figEf}. The maximum value of $S$ can be deduced as below
\bea
\frac{\partial S}{\partial E_{0}}=0,
\eea
And, therefore,
\bea
E_{0}=\frac{\pi (m^{2}+k_{\perp}^{2})}{q\ln(2)},\label{smax}
\eea
 which is equivalent to $|\beta_{k}|^2=\frac{1}{2}$. In Fig. \ref{figmf} the entanglement entropy is obtained by substituting $|\beta_{k}|^{2}=\exp(\frac{-\pi k_{\perp}^{2}}{2qE_{0}})$ in Eq. (\ref{ssfermion}), for massless particles. Large values of the electric fields correspond to large values of $|\beta_{k}|^2$ and therefore the small value of $S_{k}$. According to Eq. (\ref{ssfermion}), for fixed values of $E_{0}$, $k_{\perp}$ and $q$, the maximum value of entropy is equal to one for $m^{2}=\frac{E_{0}q\ln(2)}{\pi}-k_{\perp}^{2}$.

\section{THE SAUTER-TYPE ELECTRIC FIELD AND THE ENTANGLEMENT GENERATION}
In the previous Section, we showed that the constant electric field can generate the entanglement and worked out its variation. Now, one can explore  entanglement generation by the
pulsed electric field for scalar particles and Dirac fermions.
\subsection{ Scalar Particles }

 Eq. (\ref{equation}) can be rewritten for Sauter-type electric field along
the $z$ direction as: $E(t)=E_{0}sech^{2}(t/\tau)$ \cite{sauter} in which  $\tau$
 is the width of the electric field.
 One can choose $A_{\mu}$ as
\bea
A_{\mu}=(0,0,0,-E_{0}\tau \tanh(\frac{t}{\tau})),
\eea
The Fourier time component of the Klein-Gordon equation for the scalar particle with zero spin satisfies the equation
\bea
[\partial^{2}_{t}+\omega^{2}_{k}(t)]\phi_{k}(t)=0,\label{Etss}
\eea
where
\bea
\omega^{2}_{k}(t)=[k_{z}-qE_{0}\tau \tanh(\frac{t}{\tau})]^{2}+k^{2}_{\perp}
+m^{2}
\eea
 As before, we use the asymptotic solutions at $t_{in}=-\infty$ and $t_{out}=\infty$ in order to expand the separable '$in$' state in terms of the entangled '$out$' state.
 In the following,  the Bogoliubov coefficient which relates the asymptotic solutions to each other is used to obtain the reduced density matrix and the von-Neumann entropy for scalar particles.

Ref. \cite{gradshteyn} gives two linearly independent solutions of Eq. (\ref{Etss}):
\bea
\phi_{k}(t)= (z-1)^{i\tau\omega_{k,out}/2}z^{-i\tau\omega_{k,in}/2} [C_{1} F(a,b,c;z)\nonumber\\
+C_{2}z^{1-c}F(a-c+1,b-c+1,2-c;z)],\label{hyper}\nonumber\\
  \eea
where, $F$ is the hypergeometric function, and
\bea
z&=&\frac{1}{2}\tanh(\frac{t}{\tau})+\frac{1}{2},     \lambda=\sqrt{(qE_{0}\tau^{2})^{2}-\frac{1}{4}},\nonumber\\
 a&=&\frac{1}{2}+\frac{i}{2}(\tau\omega_{k,out}-\tau\omega_{k,in})-i\lambda,\nonumber\\
 b&=&\frac{1}{2}+\frac{i}{2}(\tau\omega_{k,out}-\tau\omega_{k,in})+i\lambda,\nonumber\\
 c&=&1-i\tau\omega_{k,in}.\label{z}
 \eea
 in which, $\omega_{in}$ and $\omega_{out}$ are the kinetic energies of the field modes at asymptotic times $t_{in}=-\infty$ and $t_{out}=\infty$
 \bea
 \omega_{k,in}&=&\sqrt{(k_{z}+qE_{0}\tau)^{2}+k^{2}_{\perp}+m^{2}},\nonumber\\
 \omega_{k,out}&=&\sqrt{(k_{z}-qE_{0}\tau)^{2}+k^{2}_{\perp}+m^{2}}.
 \eea
  As we know, the asymptotic solutions are related through the Bogoliubov coefficients. Using the properties of the hypergeometric function discussed in more detail in appendix, one can find Bogoliubov coefficients as follows
  \bea
  |\beta_{k}|^2&=&\frac{\cosh [\pi\tau(\omega_{out}-\omega_{in})]+\cosh (2\pi\lambda)}{2\sinh(\pi\tau\omega_{in})\sinh(\pi\tau\omega_{out})}\nonumber\\
  |\alpha_{k}|^2&=&\frac{\cosh [\pi\tau(\omega_{out}+\omega_{in})]+\cosh (2\pi\lambda)}{2\sinh(\pi\tau\omega_{in})\sinh(\pi\tau\omega_{out})}\label{b2a22}.
  \eea

 \noindent We can use the method described in Section II above to
expand the '$in$'- vacuum in terms of the '$out$' state and specify the coefficient $c_{n}$. 
 Below,  Eqs. (\ref{cn}-\ref{scalarentropy}) and (\ref{b2a22}) are used to obtain the von Neumann entropy for scalar particles in the  background of the Sauter-type electric field, we have
 \bea
 S&=&\log_{2}\frac{x^{\frac{x}{x-1}}}{1-x}\nonumber\\
 x&=& \frac{|\beta_{k}|^2}{|\alpha_{k}|^2}=\frac{\cosh [\pi\tau(\omega_{out}-\omega_{in})]+\cosh (2\pi\lambda)}{\cosh [\pi\tau(\omega_{out}+\omega_{in})]+\cosh (2\pi\lambda)}.\nonumber\\\label{alphaboson}
 \eea

\begin{figure}[t]
    \center
    \includegraphics[width=0.7\columnwidth]{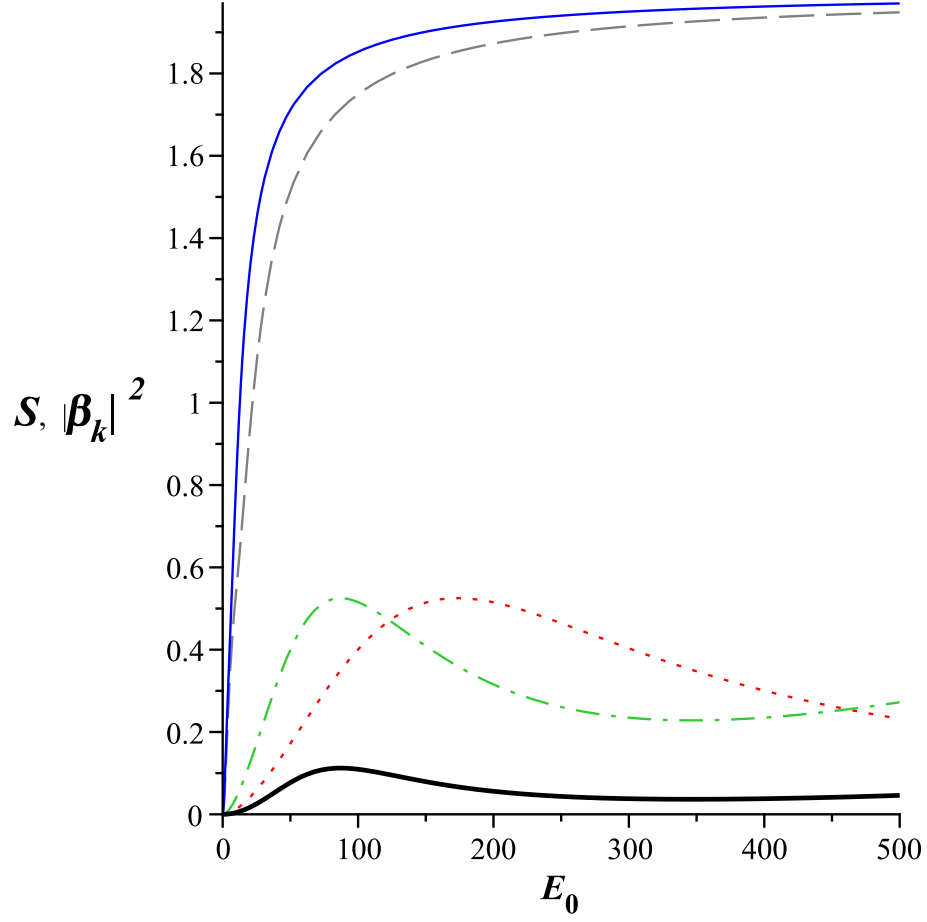}\\
    \caption{(Color online) The von Neumann entropy for bosonic modes with respect to the amplitude of a pulsed electric field for ($m, q, k_{\perp}, k_{z}=1$) and certain values of  $\tau$: $\tau=.3$  (blue solid line),  $\tau=.2$  (gray dashed line), $\tau=0.02$ ( green dash-dotted line), and  $\tau=0.01$ ( red dotted line). Also, $|\beta_{k}|^{2}$ as a function of $E_{0}$ is represented by the lower solid thick black line for $\tau=0.02$. The general behavior of $|\beta|^{2}$ is the same that of the von Neumann entropy. For small $\tau$, $|\beta_{k}|^{2}$ has a local maximum and for fixed values of $m, q, K_{\perp}, k_{z}, \tau$, the related $E_{0}$ can be obtained numerically. By substituting $(|\beta_{k}|^{2})_{max}$ in Eq. (\ref{alphaboson}), The maximum value of $S_{k}$ is obtained. The values of electric fields which maximize $|\beta_{k}|^{2}$ and $S_{k}$ are equal. For large values of $\tau$ (blue solid line and gray dashed line), the pulsed electric field tends to constant electric field and therefore, the results are coincided to the constant one. }\label{figbtE}
   \end{figure}

\begin{figure}[t]
    \center
    \includegraphics[width=0.7\columnwidth]{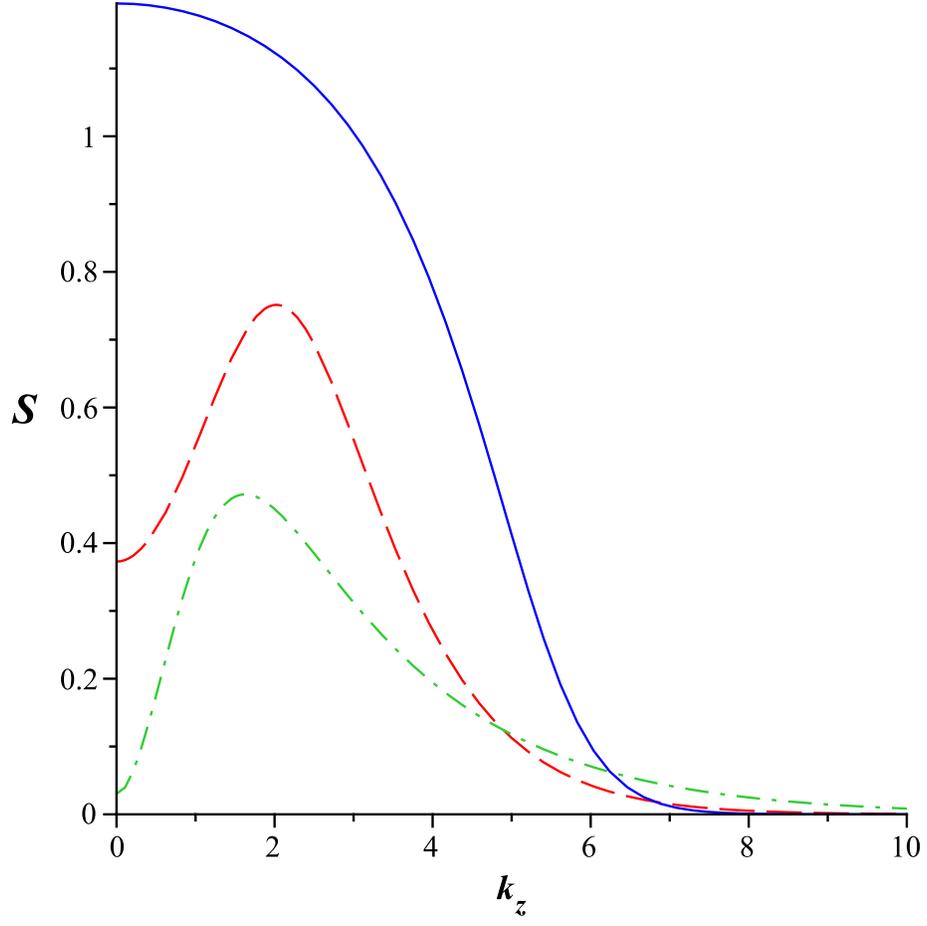}\\
    \caption{(Color online)  The von Neumann entropy for bosonic modes with respect to $k_{z}$ for ($m, q, k_{\perp}=1, E=10$) and certain values of  $\tau$: $\tau=0.5$ (blue solid line), $\tau=0.2$ ( red dashed line), and $\tau=0.1$ (green dash-dotted line). For fixed value of $(m, q, E_{0}, K_{\perp}, k_{z})$, the entanglement entropy has local maximum. The related momentum can be obtained numerically.}\label{figbtk}
   \end{figure}

\begin{figure}[b]
    \center
    \includegraphics[width=0.7\columnwidth]{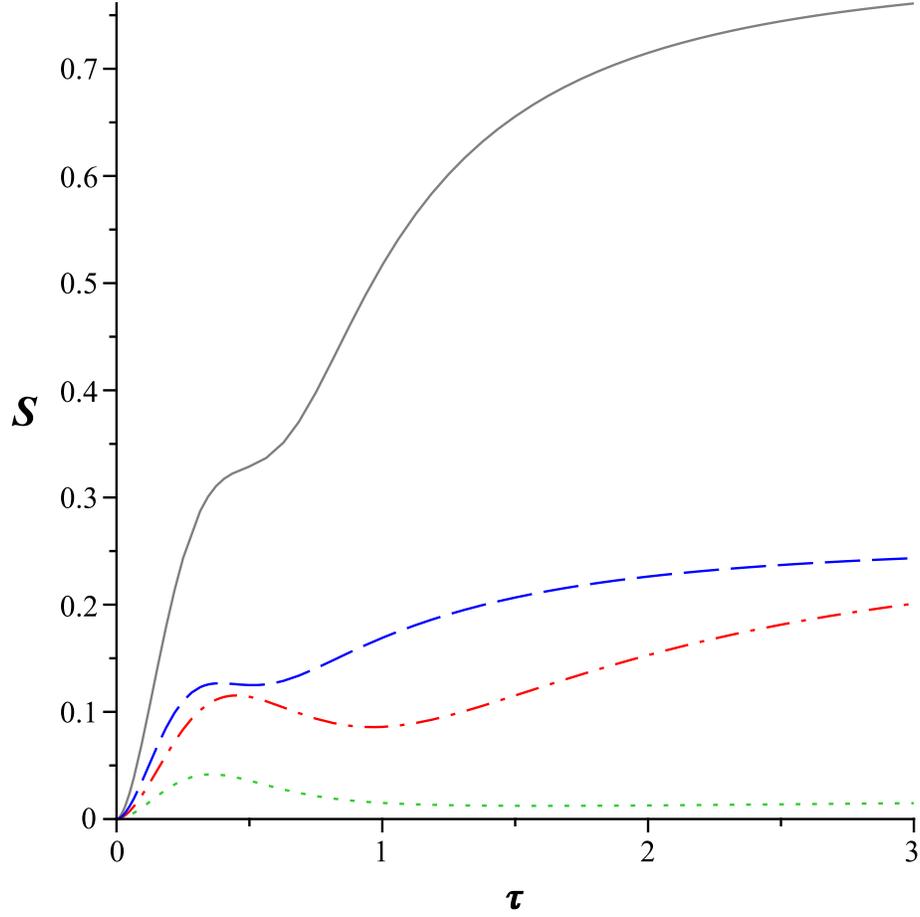}\\
    \caption{(Color online) The von Neumann entropy for bosonic modes with respect to the width of pulse $\tau$  for ($m, q, k_{z}=1$) and certain values of  $E_{0}$ and $k_{\perp}: $ $E=2,k_{\perp}=0$ ( gray solid line), $E=2,k_{\perp}=1$ (blue dashed line), $E=1,k_{\perp}=0$ (red dash-dotted line), and  $E=1,k_{\perp}=1$ (the green dotted line). When $\tau \rightarrow\infty$, the pulsed electric field tends to the constant electric field and the results are coincided to the constant one. Therefore, the value of $S_{k}$ depends on the fixed values of ($m, q, E, K_{\perp}, k_{z}$). }\label{figbtt}
   \end{figure}

Variation of the entanglement entropy for the Sauter-type electric field with respect to $E_{0}$, $k_{z}$, and $\tau$ are indicated in Figs. \ref{figbtE}-\ref{figbtt}. According to Fig. \ref{figbtE}, for a small value of $\tau$, the entanglement entropy has a local maximum while  the behavior of entanglement will be similar to that depicted in Fig \ref{figE} for large values of the same parameter.  Dependence of $S$ on the longitude momentum, $k_{z}$, is indicated in Fig. \ref{figbtk}. There is a peak for a small value of $\tau$, while this dependence will be less for large values of $\tau$. As mentioned before, for bosonic modes higher values of $|\beta_{k}|^{2}$ lead to a greater entanglement. In other words, the behavior of $S$ is similar to that of $|\beta_{k}|^{2}$.

\subsection{ Fermion Particles }
In this subsection, we explore the variation of the entanglement that is generated by Sauter-type electric field between fermionic modes. Repeating the same procedure for the constant electric field, Eqs. (\ref{dirac}-\ref{second}), one can find the spin diagonal component
of the Dirac equation for spinor QED to satisfy the equation
 \bea
[\partial^{2}_{t}&+&[k_{z}-qE_{0}\tau \tanh(\frac{t}{\tau})]^{2}+k^{2}_{\perp}+m^{2}\nonumber\\
&+& iqE_{0}\sec h^{2}(t/\tau)]\phi_{k}(t)=0,\label{Eft}
\eea
 Two linear  solutions of Eq. (\ref{Eft}) are given by
\bea
\phi_{k,s}(t)= (z-1)^{i\tau\omega_{k,out}/2}z^{-i\tau\omega_{k,in}/2} [C_{1} F(a,b,c;z)\nonumber\\
+C_{2}z^{1-c}F(a-c+1,b-c+1,2-c;z)],\label{hyperf}\nonumber\\
 \eea
 with
 \bea
z&=&\frac{1}{2}\tanh(\frac{t}{\tau})+\frac{1}{2},     \lambda=qE_{0}\tau^{2},\nonumber\\
 a&=&\frac{i}{2}(\tau\omega_{k,out}-\tau\omega_{k,in})\pm i\lambda,\nonumber\\
 b&=&1+\frac{i}{2}(\tau\omega_{k,out}-\tau\omega_{k,in})\mp i\lambda,\nonumber\\
 c&=&1-i\tau\omega_{k,in}.\label{zf}
 \eea

\begin{figure}[t]
    \center
    \includegraphics[width=0.7\columnwidth]{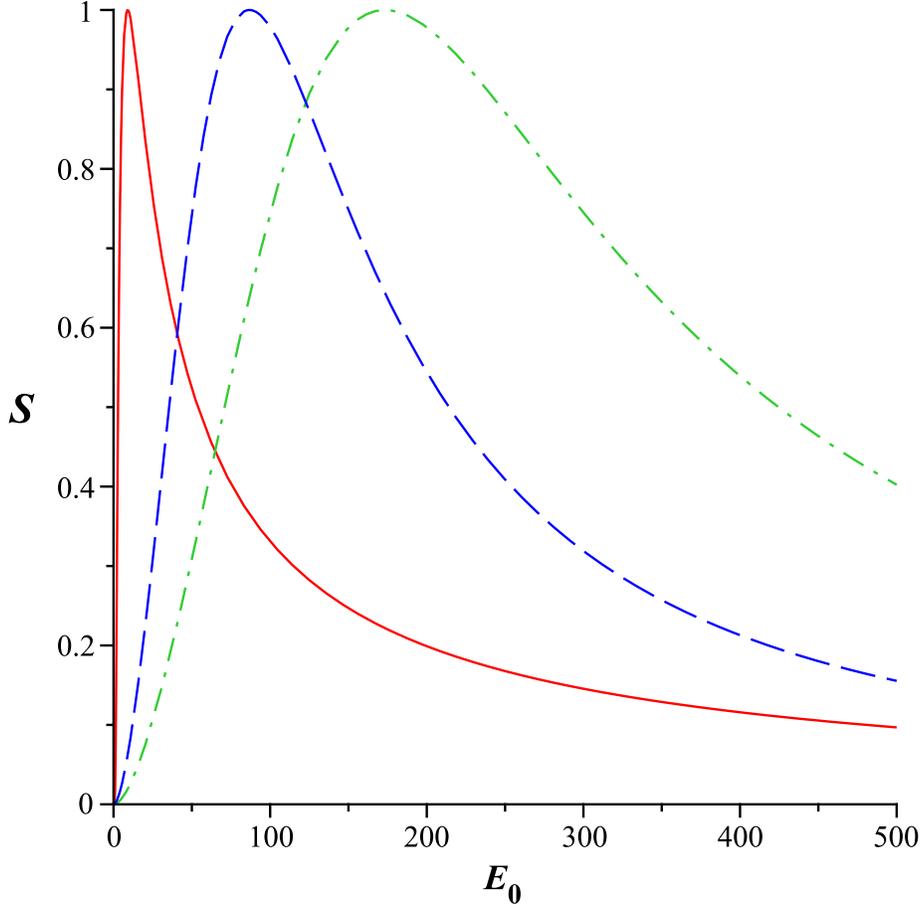}\\
    \caption{(Color online) The von Neumann entropy for fermionic modes with respect to $E_{0}$ for ($m, q, k_{z}, k_{\perp}=1$) and some values of the $\tau$, ($\tau=2$)(red solid line), ($\tau=0.02$)(blue dashed line), ($\tau=0.01$) (green dash-dotted line). According to Eq. (\ref{alphabe}), in very small or large electric fields, $|\beta_{k}|^{2}$ tends to zero or one, respectively. Therefore, according to Eq. (\ref{sssfermion}), the entropy has its minimum value, $S_{min}=0$. The maximum value of entropy $(S_{max}=1)$, for fixed values of different parameters $(m, q, k_{z}, k_{\perp}, \tau)$ occurs at a specified electric field wherein $|\beta_{k}|^{2}=1/2$. The value of this electric field can be obtained numerically.
    }\label{figftE}
   \end{figure}

\begin{figure}[b]
    \center
    \includegraphics[width=0.7\columnwidth]{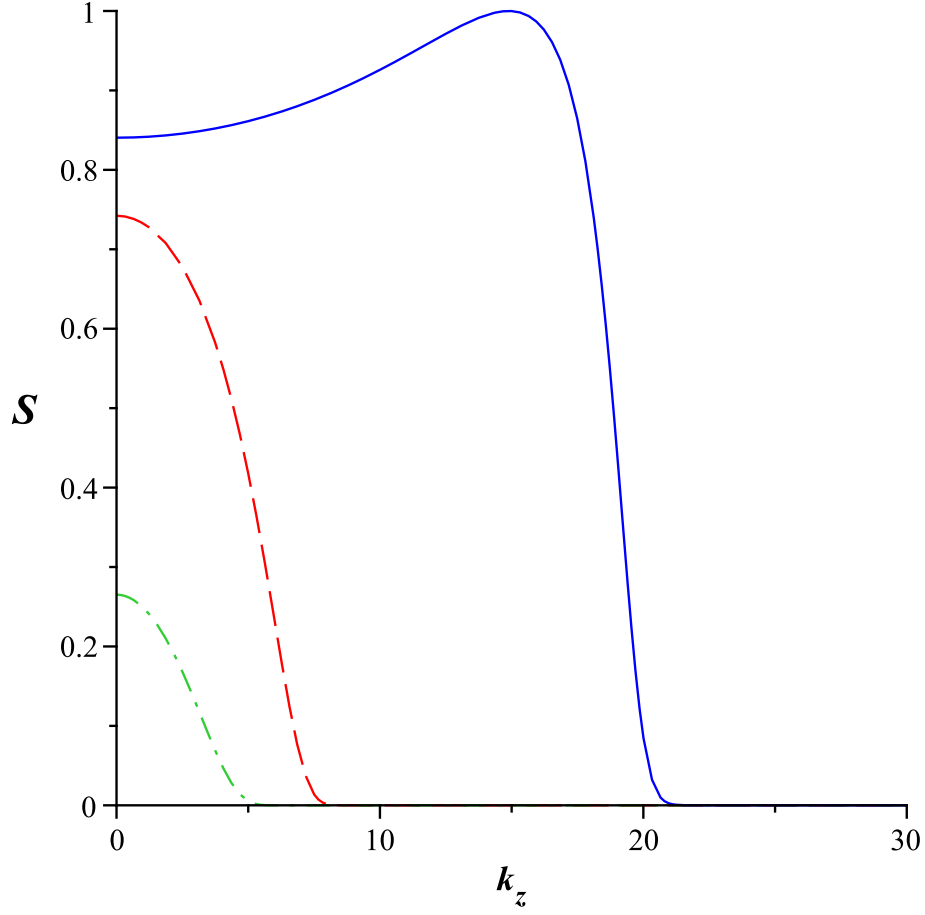}\\
    \caption{(Color online) The von Neumann entropy for fermionic modes with respect to $k_{z}$ for ($m,q,k_{\perp}=1$) and certain values of $E_{0}$ and $\tau$: $E=20,\tau=1$ (blue solid line); $E=4,\tau=2$ (red dashed line); and $E=2,\tau=3$ (green dash-dotted line). The maximum value of entropy $(S_{max}=1)$, for fixed values of different parameters ($m, q,...$), occurs wherein $|\beta_{k}|^{2}=1/2$. The related momentum can be obtained numerically. For $k_{z}=0$; $0<|\beta_{k}|^{2}<1$ and therefore $S_{k}\neq 0$. The value of $S_{k}$ is obtained by substituting $|\beta_{k}|^{2}$ in Eq. (\ref{sssfermion}). }\label{figftk}
   \end{figure}

\begin{figure}[b]
    \center
    \includegraphics[width=0.5\columnwidth]{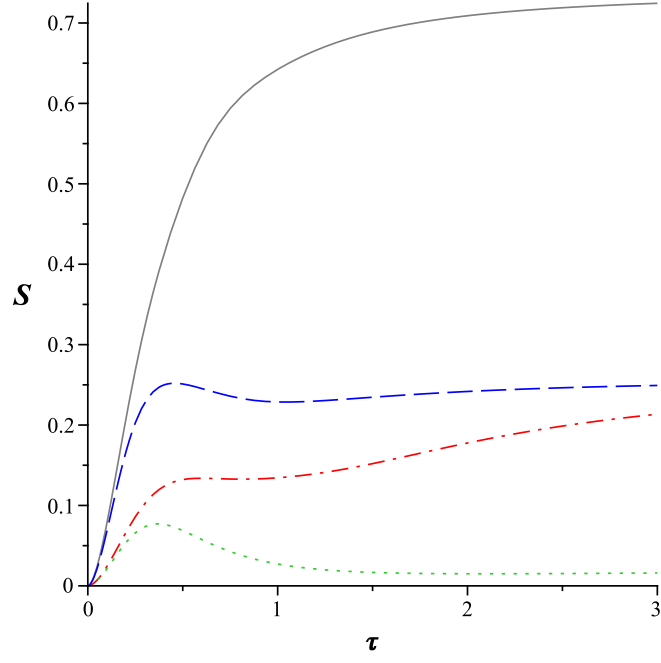}\\
    \caption{(Color online) The von Neumann entropy for fermionic modes with respect to the width of pulse $\tau$ for ($m, q, k_{z}=1$) and some certain values of  $E_{0}$ and $k_{\perp}:$  $E=2, k_{\perp}=0$ ( gray solid line);  $E=2, k_{\perp}=1$ (blue dashed line);  $E=1, k_{\perp}=0$  (red dash-dotted line); and  $E=1, k_{\perp}=1$ (green dotted line). For $\tau=0$; $|\beta_{k}|^{2}=0$ and therefore, $S_{k}=0$. When $\tau \rightarrow\infty$, the pulsed electric field tends to the constant electric field and the results are coincided to the constant one. Therefore, the value of $S_{k}$ depends on the fixed values of ($m, q, E, K_{\perp}, k_{z}$). }\label{figftt}
   \end{figure}

Using Eqs. (\ref{R},\ref{uv},\ref{innerproduct},\ref{fabb}) and the properties of the hypergeometric function, one can find Bogoliubov coefficients as follows
\bea
|\beta^{\uparrow}_{k}|^{2}&=&|\beta^{\downarrow}_{k}|^{2}=|\beta_{k}|^{2}=\frac{\cosh(2\pi\lambda)-\cosh[\pi\tau(\omega_{out}-\omega_{in})]}{2\sinh(\pi\tau\omega_{in})\sinh(\pi\tau\omega_{out})}\nonumber\\
|\alpha^{\uparrow}_{k}|^{2}&=&|\alpha^{\downarrow}_{k}|^{2}=|\alpha_{k}|^{2}=\frac{\cosh[\pi\tau(\omega_{out}+\omega_{in})]-\cosh(2\pi\lambda)}{2\sinh(\pi\tau\omega_{in})\sinh(\pi\tau\omega_{out})}\nonumber\\
\label{alphabe}
\eea
where, $\uparrow(\downarrow)$ indicate the up (down) spins.
Exploiting the method described in section II, we may expand the '$in$'- vacuum in terms of the '$out$' states according to Eq. (\ref{fermionstate}). The coefficient ($c_{0},...,c_{3}$)  given by Eq. (\ref{cnf}), may now be used to obtain the reduced density matrix and, thereby, the average von-Neumann entropy Eq. (\ref{sfermion}). In fact, the relations (\ref{abspinor}-\ref{ro}) remain valid. Using these relations, the entropy of von Neumann is obtained by:
\bea
S=-|\alpha_{k}|^{2}\log_{2}(|\alpha_{k}|^{2})-|\beta_{k}|^{2}\log_{2}(|\beta_{k}|^{2})\label{sssfermion},
\eea
where, $|\alpha|^2$ and $|\beta|^2$  are specified by (\ref{alphabe}).
The variations of the entanglement with respect to $E_{0}$, $\tau$ and $k_{z}$ are indicated in Figs. \ref{figftE}-\ref{figftt}. The behavior of $S$ with respect to $m$ and $k_{\perp}$ for the pulsed and constant electric fields will be is the same. Eq. (\ref{sssfermion}) indicates that the maximum value of the entanglement entropy occurs at $|\beta|^2=\frac{1}{2}$. The minimum value of $S_{k}$ also occurs at $|\beta|^2$ is equal to one or zero. 

When $\tau\rightarrow\infty$, the pulsed electric field $E(t)=E_{0} sech^{2}(t/\tau)$ tends to the constant electric field $E_{0}$. Also, in this limit, the Bogoliubov coefficients for bosonic and fermionic modes of Eqs. (\ref{b2a22}) and (\ref{alphabe}) tend to be as in Eqs. (\ref{alpha}) and (\ref{fab}). Since the generated entanglement is obtained in terms of Bogoliubov coefficients, the behavior of $S$ generated by the pulsed electric field can be observed to tend to the constant electric field for large values of $\tau$.

\section{ CONCLUSION}
We applied the Schwinger pair production theory to constant and pulsed electric fields on a Minkowski spacetime to demonstrate that the background electric field can generate the entanglement. We worked out the entanglement entropy for scalar particles and Dirac fermions  created by the background electric field. The behavior of the entanglement was also depicted in figures with respect to different parameters.

For a constant electric field, the entanglement generated for boson modes as a function of $E_{0}$ was found to be an increasing function which tended to a specific constant value in the limit of sufficiently large fields (Fig. \ref{figE}) but that it monotonically decreased with respect to $m$ and $(k_{\perp})$ ( Fig. \ref{figm}). For the fermionic mode, however, it was found to be very different. Optimal values of $E_{0}$, $m$ and $k_{\perp}$ were observed for which the entanglement entropy would be maximized (Figs. \ref{figEf} and \ref{figmf}).

In the case of a pulsed electric field, the behavior of the entanglement generated by bosonic and fermionic modes with respect to $m$ and $k_{\perp}$ was observed to be similar to that in the case of the constant electric field. However, for small values of $\tau$, the bosonic entanglement as a function of $E_{0}$ was seen to have a local maximum (Fig. \ref{figbtE}). High power laser pulses make a good candidate for  generating entangled states experimentally.

 The authors intend to explore the generation of entanglement by other forms of electric and magnetic fields in future. As another interesting area of
 research is to study entanglement generation by background electromagnetic fields at finite temperature.

\section{Appendix A:Hypergeometric function and Bogoliubov coefficients}
 The hypergeometric function, $F(a,b,c,z)$, appears in $24$ different forms. These solutions are called Kummer's series and can be arranged in six sets such that the four series belonging to each set represent the same function. Any three of these are connected by a linear relation with constant coefficients \cite{gradshteyn}. We choose appropriate solutions among the six sets for $t=+\infty$ and $t=-\infty$. As we know, the asymptotic solutions are related through the Bogoliubov coefficients
 \bea
 \phi_{k,in}= \alpha_{k}\phi_{k,out}+\beta_{k}\phi^{*}_{k,out}\label{bg}.
 \eea

  Using the behavior of the hypergeometric function, we can find the Bogoliubov coefficients. According to (\ref{z}), the value of $z$ at $t=-\infty(+\infty)$ is $0(1)$ . Therefore, the  hypergeometric functions can be appropriately expressed in terms of $z$ and $1-z$.

 Using the linear relations between the hypergeometric functions \cite{gradshteyn},
 \begin{widetext}
 \bea
 F(a,b,c;z)=\lambda_{11} F(a,b,a+b-c+1;1-z) +\lambda_{12}(1-z)^{c-a-b} F(c-a,c-b,c-a-b+1;1-z)\label{f1}
 \eea
 and
  \bea
z^{1-c}F(a-c+1,b-c+1,2-c;z)&=&\lambda_{21} F(a,b,a+b-c+1;1-z)\nonumber\\
&+&\lambda_{22}(1-z)^{c-a-b}F(c-a,c-b,c-a-b+1;1-z),\label{f2}
 \eea
 where
 \bea
 \lambda_{11}&=&\frac{\Gamma(c)\Gamma(c-a-b)}{\Gamma(c-a)\Gamma(c-b)},\ \ \lambda_{21} =\frac{\Gamma(2-c)\Gamma(c-a-b)}{\Gamma(1-a)\Gamma(1-b)}\nonumber\\
\lambda_{12}&=&\frac{\Gamma(c)\Gamma(a+b-c)}{\Gamma(a)\Gamma(b)},\ \ \lambda_{22} =\frac{\Gamma(2-c)\Gamma(a+b-c)}{\Gamma(a-c+1)\Gamma(b-c+1)},\label{f3}
 \eea
 \end{widetext}
  one can expand the sets of solutions at $t_{in}= − \infty$ in terms of the sets of solutions at $t_{out}=+\infty$.
  Using Eqs. (\ref{bg}-\ref{f3}), we have
   \bea
 \frac{|\beta_{k}|^2}{|\alpha_{k}|^2}=\frac{| \lambda_{12}|^2}{| \lambda_{11}|^2}=\frac{| \lambda_{22}|^2}{| \lambda_{21}|^2}\label{b2a2}.
  \eea


\end{document}